\documentclass{webofc}
\usepackage[varg]{txfonts}   % Web of Conferences font
\begin{document}
\title{Thermal-model-based characterization of heavy-ion-collision systems at chemical freeze-out}

\author{\firstname{Jamie M.} \lastname{Karthein}\inst{1,2}\fnsep\thanks{\email{jmstafford@uh.edu}} \and
        \firstname{Paolo} \lastname{Alba}\inst{3}
\and
        \firstname{Valentina} \lastname{ Mantovani-Sarti}\inst{4}
        \and
        \firstname{Jacquelyn} \lastname{ Noronha-Hostler}\inst{5}
         \and
        \firstname{Paolo} \lastname{ Parotto}\inst{6}
         \and
        \firstname{Israel} \lastname{ Portillo-Vazquez}\inst{1}
         \and
        \firstname{Volodymyr} \lastname{  Vovchenko}\inst{2}
         \and
        \firstname{Volker} \lastname{ Koch}\inst{2}
         \and
        \firstname{Claudia} \lastname{ Ratti}\inst{1}
}

\institute{Department of Physics, University of Houston, Houston, Texas 77204, USA
\and
          Nuclear Science Division, Lawrence Berkeley National Laboratory, Berkeley, CA 94720, USA
\and
           Lucht Probst Assoc., GmbH, Frankfurt 60311, DE 
\and
           Department of Physics, Technical University Munich, Munich 80333, DE
\and
           Illinois Center for Advanced Studies of the Universe, Department of Physics, University of Illinois at Urbana-Champaign, Urbana, IL 61801, USA
\and
           Department of Physics,  University of Wuppertal, Wuppertal 42119, DE
          }

\abstract{
  We investigate the chemical freeze-out in heavy-ion collisions (HICs) and the impact of the hadronic spectrum on thermal model analyses \cite{Alba:2020jir,Bellwied:2018tkc}. Detailed knowledge of the hadronic spectrum is still an open question, which has phenomenological consequences on the study of HICs. By varying the number of resonances included in Hadron Resonance Gas (HRG) Model calculations, we can shed light on which particles may be produced. Furthermore, we study the influence of the number of states on the so-called two flavor freeze-out scenario, in which strange and light particles can freeze-out separately. We consider results for the chemical freeze-out parameters obtained from thermal model fits and from calculating net-particle fluctuations. We will show the effect of using one global temperature to fit all particles and alternatively, allowing particles with and without strange quarks to freeze-out separately.
}
\maketitle
\section{Introduction}
\label{intro}
Ultra-relativistic heavy-ion collisions provide an opportunity to study the  transition from ordinary hadronic matter to the deconfined Quark-Gluon Plasma. Currently, calculations of the fundamental theory with Lattice Quantum Chromodynamics have shown there exists a crossover transition at low baryon chemical potential around T $\sim$ 156 MeV \cite{Aoki:2006br}. During this transition, the system traverses through chemical and kinetic freeze-out stages. The location of these freeze-out points in the phase diagram is an important characterization of heavy-ion collisions. Statistical and thermal models can be used to determine those freeze-out parameters, $T_f$, $\mu_{B,f}$ (and $V_f$), at chemical freeze-out \cite{Andronic:2017pug,Vovchenko:2019pjl,Alba:2014eba}. Thermal fit analyses utilize the $\chi^2$ minimization procedure for a fit to experimental data of various particle species \cite{Bellini:2018khg,Adamczyk:2017iwn,Andronic:2017pug}.  Alternatively, one can determine the freeze-out parameters by utilizing net-charge fluctuations \cite{Alba:2014eba}. This method allows for the determination of freeze-out parameters for different particle species separately, for example for light particles and kaons \cite{Alba:2014eba,Bellwied:2018tkc}. In these proceedings, freeze-out parameters are analyzed both via thermal fits and net-particle fluctuations.
\section{HRG Model}
\label{sec-1}
This study of the chemical freeze-out stage in heavy-ion collisions is performed within the Hadron Resonance Gas (HRG) model framework. The interacting hadrons are well-described by an ideal gas of hadrons and resonances \cite{Bazavov:2012jq, Borsanyi:2010bp, Vovchenko:2016rkn},
\begin{align*}
\centering
\frac{P}{T^4} = \frac{1}{VT^3} &\sum_i \ln Z_i (T, V, \vec{\mu}),
\\
\ln  Z^{M/B}_i  = \mp \frac{V \, d_i}{(2\pi)^3} \int d^3 k \, &\ln  \left( 1  \mp \, exp \left[ - \left(\epsilon_i - \mu_a X_a^i \right)/T \right] \right)
\end{align*}
where the index i runs over all the particles included in the HRG model, the energy for the single particle reads $\epsilon_i = \sqrt{k^2 + m^2_i}$, the conserved charges $X_i$ are $ B_i, S_i, Q_i $, $d_i$ and $m_i$ represent the degeneracy and the mass for the $i$th particle and V is the volume.
\subsection{Thermal model fits}
\label{sec-2.1}
\begin{figure*}
    \centering
    \includegraphics[width=0.32\textwidth]{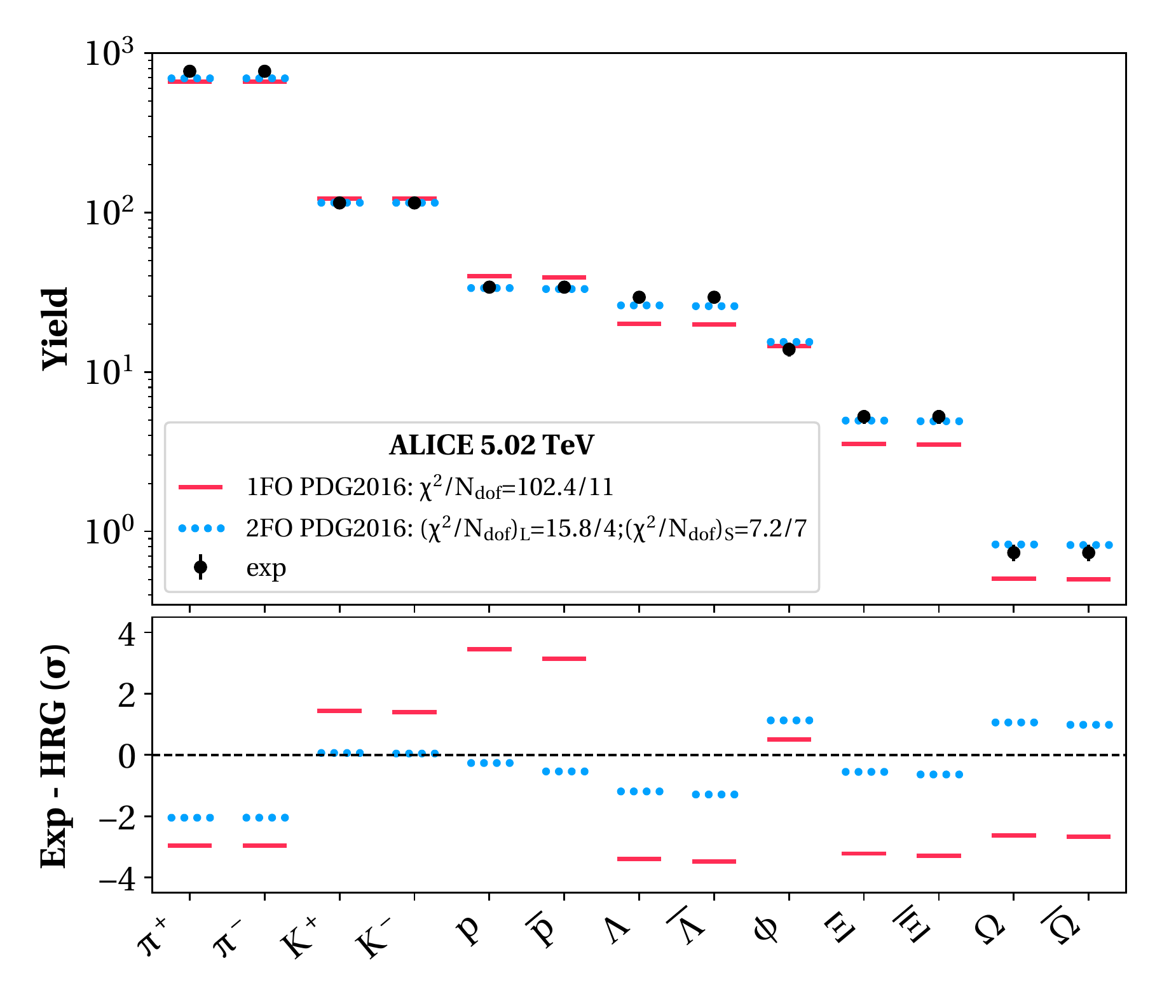}
    \includegraphics[width=0.32\textwidth]{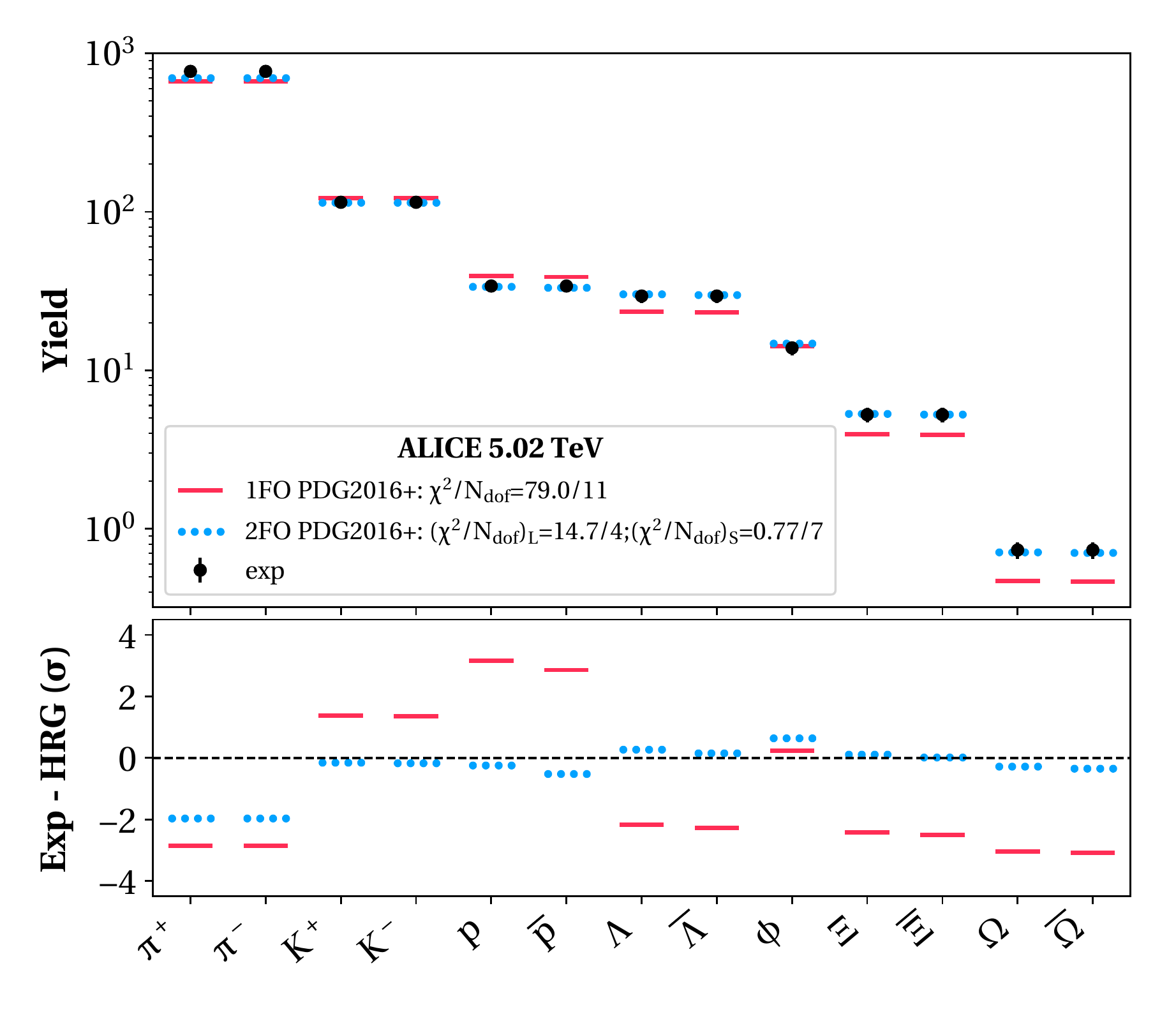}
    \includegraphics[width=0.32\textwidth]{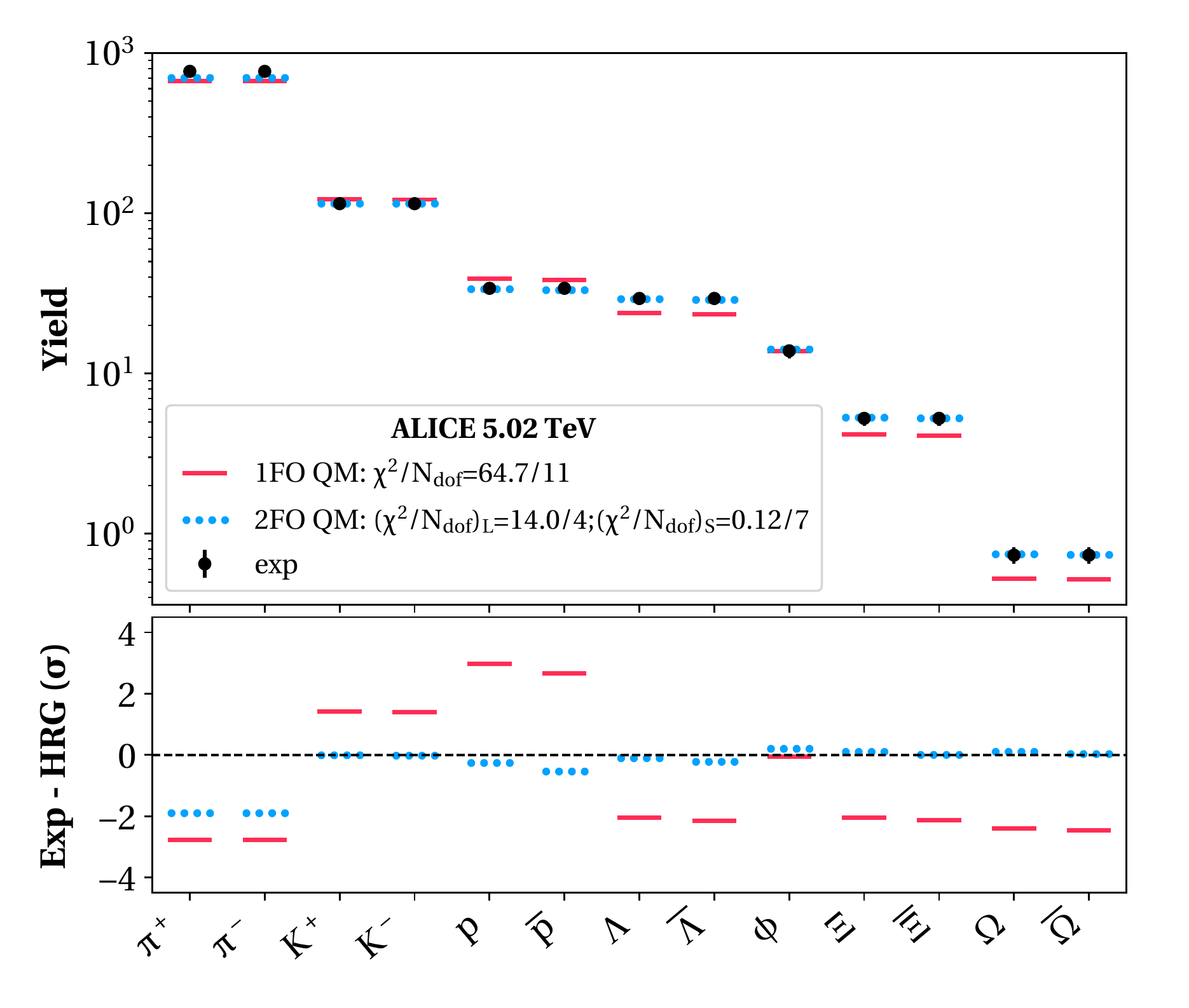}
    \caption{ALICE PbPb $\sqrt{s_{NN}}=5.02$ GeV data for particle yields in $0-10\%$ collisions, in comparison to HRG model calculations with different resonance lists; deviations in units of experimental errors $\sigma$ are 
shown below each panel.}
    \label{fig:fits}
\end{figure*}

The HRG model has been widely employed to compare data on particle
production for a wide range of energies~\cite{Becattini:2005xt,Becattini:2012xb,Cleymans:2005xv,Torrieri:2006yb,Andronic:2011yq,BraunMunzinger:2003zd,Andronic:2005yp}.
Produced particle yields are calculated by adding the contribution from resonances to the primordial yields. In these proceedings, we focus on the ALICE data at $\sqrt{s_{\mathrm{NN}}}=5.02$ TeV 
and $0-10\%$ centrality and perform thermal fits of the particle yields ~\cite{ABELEV:2013zaa}. In Fig. \ref{fig:fits} and Table \ref{tab:ALICE}, we show the fits for the yields for the PDG2016, PDG2016+, and Quark Model hadronic lists (described in Ref. \cite{Alba:2020jir}) and extract the thermal parameters $(T,~\mu_B,~V)$ by using the thermal fit package in FIST \cite{Vovchenko:2019pjl}. The FIST software allows users to choose their own particle lists, as well as data sets, 
in the fit.
\begin{table*}
\centering
\begin{tabular}{| c | c | c | c | c |}
\hline
 & & T [MeV] & Volume [$\text{fm}^3$]  & $\chi^2$ \\
 \hline
PDG2016  & Single FO &152.2 $\pm$ 1.6 & 5797 $\pm$ 523 & 102.4/12 \\
\hline
& Light  &143.2 $\pm$ 1.8 & 9096 $\pm$ 897 & 15.7/4  \\
\hline
 & Strange  &169.3 $\pm$ 2.5 & 5797 $\pm$ 523 & 7.7/8 \\
\hline
PDG2016+  & Single FO &150.4 $\pm$ 1.5 & 6239 $\pm$ 539 & 79.0/12 \\
\hline
& Light &142.5 $\pm$ 1.7 & 6239 $\pm$ 539 & 14.7/4  \\
\hline
& Strange &164.5 $\pm$ 2.3 & 3087 $\pm$ 367 & 1.6/8 \\
\hline
QM  & Single FO &147.9 $\pm$ 1.3 & 6852 $\pm$ 558 & 64.7/12 \\
\hline
& Light &140.9 $\pm$ 1.6 & 9961 $\pm$ 927 & 13.9/4  \\
\hline
 & Strange &158.2 $\pm$ 1.9 & 3949 $\pm$ 419 & 0.78/8 \\
\hline
\end{tabular}
\caption{Fit parameters for the two chemical freeze-out scenarios obtained from fits to total yields of ALICE data~\cite{Abelev:2013vea,ABELEV:2013zaa,Abelev:2013xaa,Abelev:2014uua} for $0-10\%$ centrality in PbPb collisions at $\sqrt{s_{NN}}=5.02$ TeV, using different lists. Here, we fix the baryon chemical potential to a value 1 MeV.}
\label{tab:ALICE}
\end{table*}
\subsection{Net-particle fluctuations}
\label{sec-2.2}
\begin{figure*}
    \centering
\includegraphics[width=0.5\textwidth]{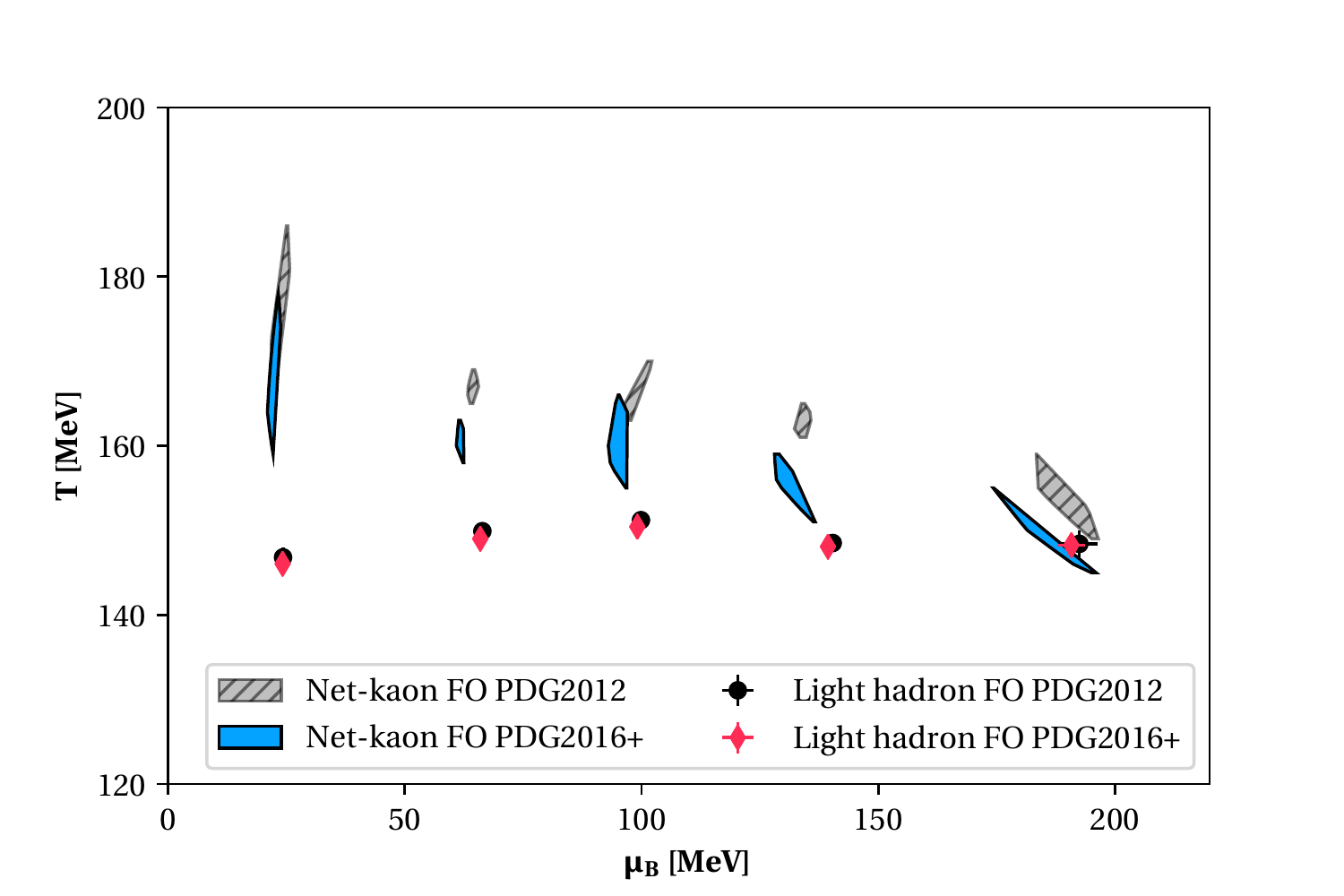}\includegraphics[width=0.42\textwidth]{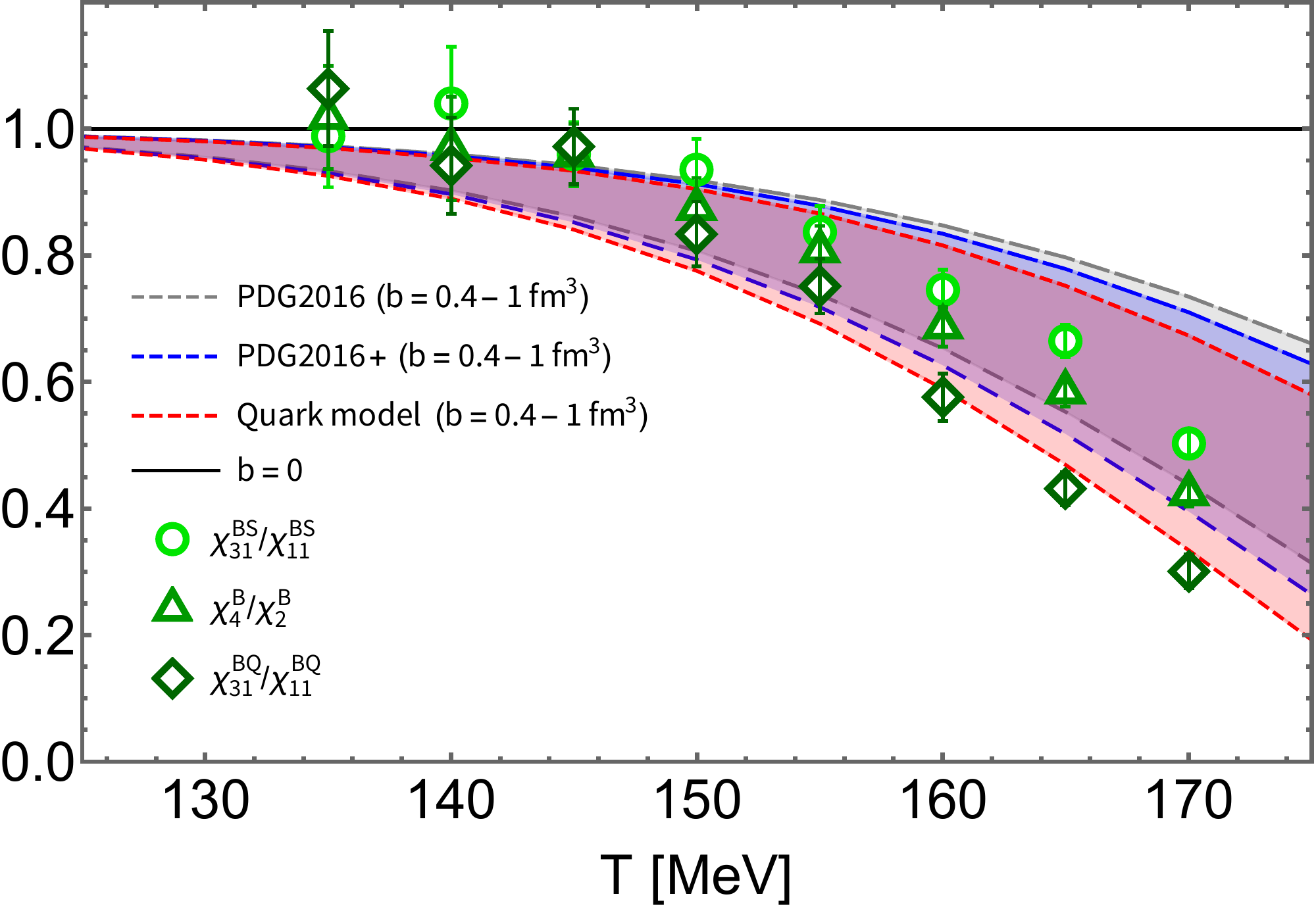}
    \caption{Left: Freeze-out parameters from light hadron and net-kaon fluctuations following the analysis from Ref. \cite{Bellwied:2018tkc} with different particle lists. Right: Fourth-to-second order susceptibilities predicted to be equal in the EV-HRG model. Lattice data from Ref. \cite{Borsanyi:2018grb} are shown in green with error bars, while the calculations within the EV-HRG model are shown as bands for a range of excluded volume parameter, $b$, and for different hadronic lists. The ideal HRG result corresponds to the line at unity.}
    \label{fig:flucts}
\end{figure*}

For the determination of freeze-out parameters from fluctuations, we employ the list PDG2016+ to perform the analysis of net-proton and net-charge fluctuations from  Ref.~\cite{Alba:2014eba}, as well as the analysis of net-kaon fluctuations from Ref.~\cite{Bellwied:2018tkc} and compare to the original results. Originally, the analysis was carried out with an older PDG list, which we indicate as PDG2012. In order to determine both $T$ and $\mu_B$ at chemical freeze-out, in general, one needs to fit two experimental quantities. However, due to the larger experimental error bars on higher order fluctuations compared to the lowest order moments, in Ref.~\cite{Bellwied:2018tkc} we obtained the freeze-out parameters as follows. Starting from the net-proton and net-charge freeze-out parameters obtained in Ref.~\cite{Alba:2014eba}, we followed the Lattice QCD isentropic trajectories determined in Ref.~\cite{Gunther:2016vcp} via a Taylor-expanded equation of state.
% These were constructed by first calculating the entropy-per-baryon ratio at each freeze-out point for the different collision energies from Ref.~\cite{Alba:2014eba}, and then following the path that conserves $S/n_B$. 
The ratio $\chi^K_1/\chi^K_2$ was then calculated along these trajectories and the freeze-out points were determined by the overlap with the experimental results, which are shown in Fig.~\ref{fig:flucts} in gray, while the red points correspond to the net-proton and net-charge freeze-out points from Ref.~\cite{Alba:2014eba}. We see that this separation in temperature for the kaons and light hadrons persists for the PDG2016+ hadronic list with many more states. Furthermore, as shown in Fig. \ref{fig:flucts}, we also find an indication of flavor-dependent freeze-out in the excluded volume HRG model when considering susceptibilities \cite{Karthein:2021cmb}. Around $T_c \sim 155$ MeV, we see a flavor dependence on the excluded volume parameter, $b$, namely a smaller excluded volume for strange baryons than non-strange baryons.
\section{Conclusions}
We have presented results on the freeze-out parameters from two different thermal analysis techniques with fits of particle yields and net-charge fluctuations. We have seen that there are two separate chemical freeze-out temperatures in heavy-ion collisions for light and strange particles. We see a better description of the experimental data when utilizing the two flavor freeze-out scenario in the thermal fits. In the case of the fluctuations analysis, we see a clear separation between the light and strange freeze-out temperatures in the phase diagram for the highest collision energies at RHIC.

\section*{Acknowledgements}
This material is based upon work supported by the National Science Foundation under grant no. PHY1654219, by the U.S. Department of Energy, Office of Science, Office of Nuclear Physics, under contract number 
DE-AC02-05CH11231231, and within the framework of the Beam Energy Scan Theory (BEST) Collaboration.   
J.N.H. acknowledges support from the Alfred
P Sloan Fellowship and the US-DOE Nuclear Science
Grant No. DE-SC0020633. P.P. also acknowledges support by the DFG grant SFB/TR55.
V.V. acknowledges the support via the
Feodor Lynen program of the Alexander von Humboldt
foundation. 

\bibliography{all.bib}

\end{document}